\input{aipcheck}

\documentclass[
    ,final            % use final for the camera ready runs
  ]
  {aipproc}

\layoutstyle{8x11single}

\begin{document}

\title{The BiPo detector for ultralow radioactivity measurements}


\author{Mathieu BONGRAND}{
  address={LAL, Universit\'e Paris-Sud 11, CNRS/IN2P3, Orsay, France.}
}

\author{the SuperNEMO Collaboration}{}

\begin{abstract}
The development of BiPo detectors is dedicated to the measurement of extremely high radiopurity in $^{208}$Tl and $^{214}$Bi  for the SuperNEMO double beta decay source foils. A modular prototype, called BiPo-1, with 0.8~m$^2$ of sensitive surface area, has been running in the Modane Underground Laboratory since February, 2008. The goal of BiPo-1 is to measure the different components of the background and in particular the surface radiopurity of the plastic scintillators that make up the detector. The first phase of data collection has been dedicated to the measurement of the radiopurity in $^{208}$Tl. After more than one year of background measurement, a surface activity of the scintillators of $\mathcal{A}$($^{208}$Tl)~$=$~1.5~$\mu$Bq/m$^2$ is reported here. 

Given this level of background, a larger BiPo3 detector having 3.25~m$^2$ of active surface area, will able to qualify the radiopurity of the SuperNEMO selenium double beta decay foils with the required sensitivity of $\mathcal{A}$($^{208}$Tl)~$<$~3-4~$\mu$Bq/kg (90\% C.L.) with a six month measurement. This detector is actually under construction and will be installed in the Canfranc Underground Laboratory mid 2011.
\end{abstract}

\maketitle

%%%%%%%%%%%%%%%%%%%%%%%%%%%%%%%%%%%%%%%%%%%%
%% MAINMATTER
%%%%%%%%%%%%%%%%%%%%%%%%%%%%%%%%%%%%%%%%%%%%

\section*{Introduction}
\label{sec:introduction}

The search for neutrinoless double beta decay ($\beta\beta 0\nu$),  is a major challenge in particle physics. The observation of this decay, a process beyond the Standard Model which violates lepton number by two units, would be an experimental proof that the neutrino is a Majorana particle, i.e. identical to its antiparticle. The NEMO-3 detector~\cite{nemo3a,nemo3b,nemo3c,nemo3d,nemo3e,nemo3f,nemo3g} has been running since 2003 in the Modane Underground Laboratory and is devoted to the search of $\beta\beta 0\nu$ decay. NEMO-3 uses a combination of a tracking detector and calorimeter for the direct detection of the two decay electrons. NEMO-3 measures 10~kg of isotopes in the form of very thin radiopure foils (30-60~mg/cm$^2$) with a total surface area of 20~m$^2$. The NEMO collaboration is developing the SuperNEMO detector~\cite{supernemoa,supernemob} which is the next generation of NEMO detectors. SuperNEMO will be able to measure 100~kg of $^{82}$Se with a sensitivity of $T_{1/2}(\beta\beta 0\nu)>10^{26}$~years which will be about two orders of magnitude higher than with NEMO-3. The isotope will be in the form of thin foils (40~mg/cm$^2$) that have a total surface area of 250~m$^2$.

One of the troublesome sources of background for SuperNEMO is a possible contamination inside the source foils of $^{208}$Tl (Q$_{\beta}$~=~4.99~MeV) and $^{214}$Bi (Q$_{\beta}$~=~3.27~MeV) produced from the decay chains of $^{232}$Th and $^{238}$U respectively. In order to achieve the desired SuperNEMO sensitivity, the required radiopurities of the SuperNEMO double beta decay foils are $\mathcal{A}$($^{208}$Tl)~$<$~2~$\mu$Bq/kg and $\mathcal{A}$($^{214}$Bi)~$<$~10~$\mu$Bq/kg. A first radiopurity measurement of purified selenium samples can be performed by $\gamma$ spectrometry with ultra low background HPGe (High Purity Germanium) detectors. Nevertheless, the best detection limit that can be reached with this technique for $^{208}$Tl is around 50~$\mu$Bq/kg, which is one order of magnitude less sensitive that the required value. In order to achieve the required  sensitivity for SuperNEMO, a R$\&$D program has been performed to develop the BiPo detector dedicated to the measurement of  ultra-low levels of contamination in $^{208}$Tl and $^{214}$Bi in the foils of SuperNEMO. The final BiPo detector must be able to measure a large area (few m$^2$) of SuperNEMO double beta decay foils (12~m$^2$ per module). The measurement must not exceed a few months. Moreover, it must be able to qualify the radiopurity of the foils in their final form,  before installation into the SuperNEMO detector. The BiPo detector can also measure the surface radiopurity of other materials.

A BiPo prototype, called BiPo-1, with an active surface of 0.8~m$^2$, has been built to validate the measurement technique, and to measure the different components of the background. BiPo-1  has been running since May 2008 in the Modane Underground Laboratory. The first phase of data collection with BiPo-1 ran until June 2009 and was dedicated to the background measurement of $^{212}$Bi ($^{208}$Tl). In a second phase, new electronics was installed in July 2009 in order to measure simultaneously the $^{212}$Bi and $^{214}$Bi background. In this article, the results of $^{212}$Bi ($^{208}$Tl) measurement from the data taken only during the first phase, are presented. The $^{214}$Bi background measurement was not conclusive in this phase because of the radon background, some radon protection improvements are currently tested with BiPo-1.

The experimental principle of the BiPo detector and the possible components of observable backgrounds are described in sections \ref{sec:principle} and \ref{sec:backgrounds}. The experimental validation of the BiPo technique, the results of the background measurements in the BiPo-1 detector and the extrapolated sensitivity to a larger BiPo-3 detector are presented in section~\ref{sec:bipo3}.

\section{Measurement principle of the BiPo detector}
\label{sec:principle}

\begin{figure}[htb]
  \centering
  \includegraphics[scale=0.35]{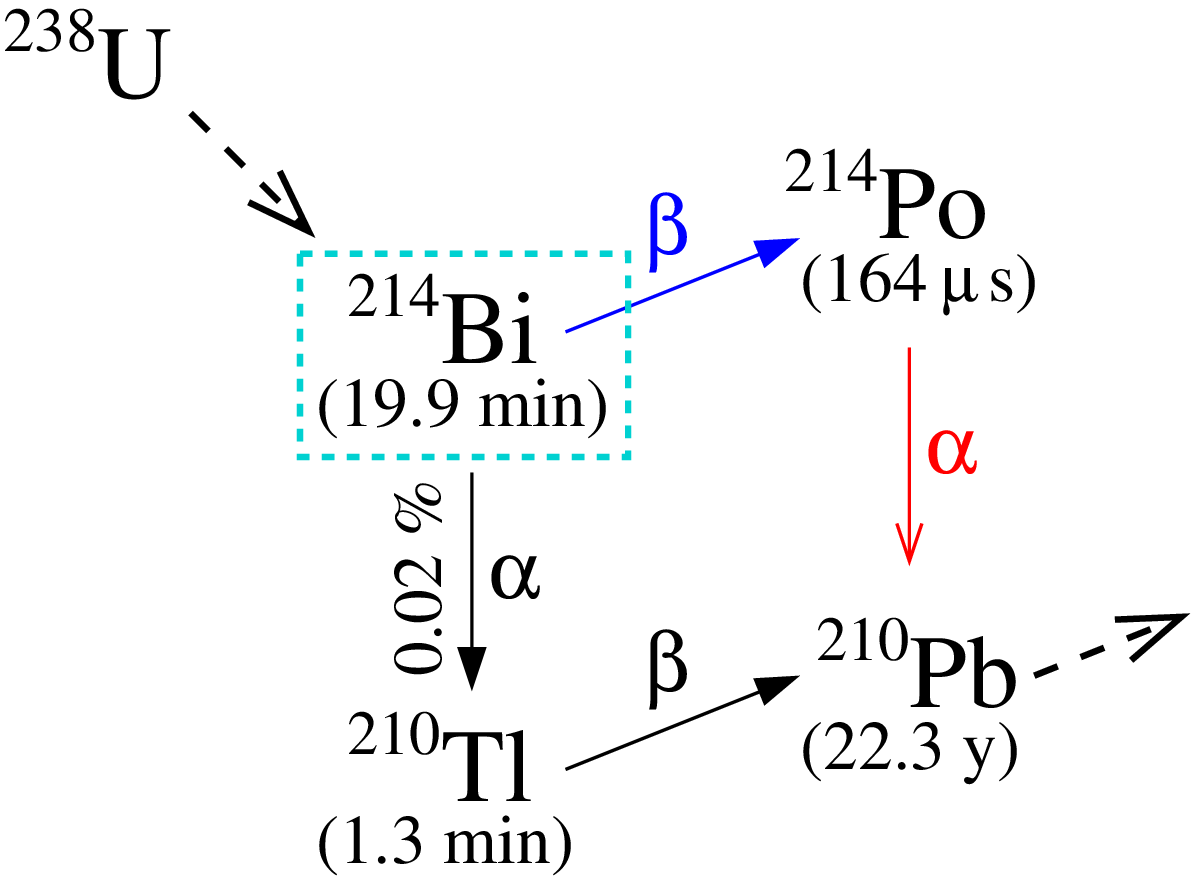}
  \hspace{1.5cm}
  \includegraphics[scale=0.35]{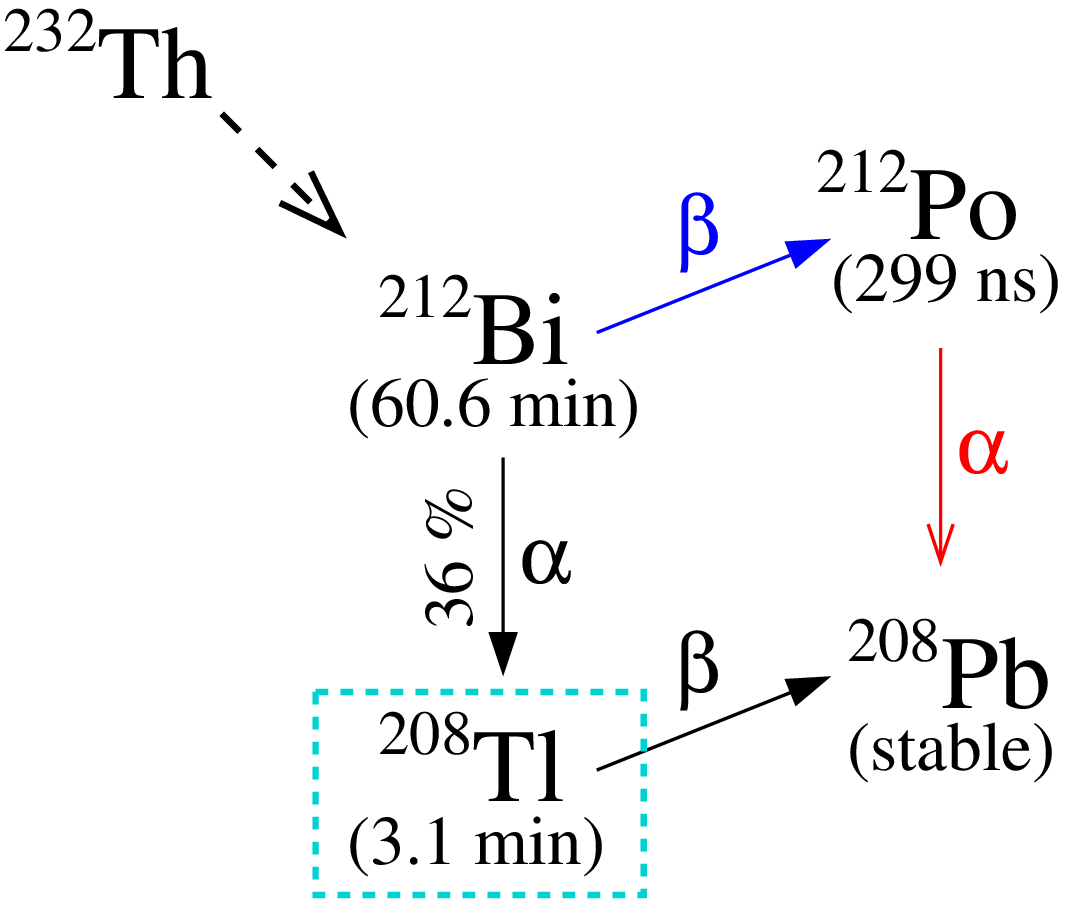}
  \caption{BiPo processes for $^{214}$Bi and $^{208}$Tl measurement.}
  \label{fig:bipo-proc}
\end{figure}

In order to measure $^{208}$Tl and $^{214}$Bi contaminations, the underlying concept of the BiPo detector is to detect with organic plastic scintillators the so-called BiPo process, which corresponds to the detection of an electron followed by a delayed $\alpha$ particle. The $^{214}$Bi isotope is a ($\beta$,$\gamma$) emitter (Q$_{\beta}$~=~3.27~MeV) decaying to $^{214}$Po, which is an $\alpha$ emitter with a half-life of 164~$\mu$s. The $^{208}$Tl isotope is measured by detecting its parent, the $^{212}$Bi. Here $^{212}$Bi decays with a branching ratio of 64\% via a $\beta$ emission (Q$_{\beta}$~=~2.25~MeV) towards the daughter nucleus $^{212}$Po which is a pure $\alpha$ emitter (8.78~MeV) with a short half-life of 300~ns (Figure~\ref{fig:bipo-proc}).

\begin{figure}[htb]
  \centering
  \includegraphics[scale=0.65]{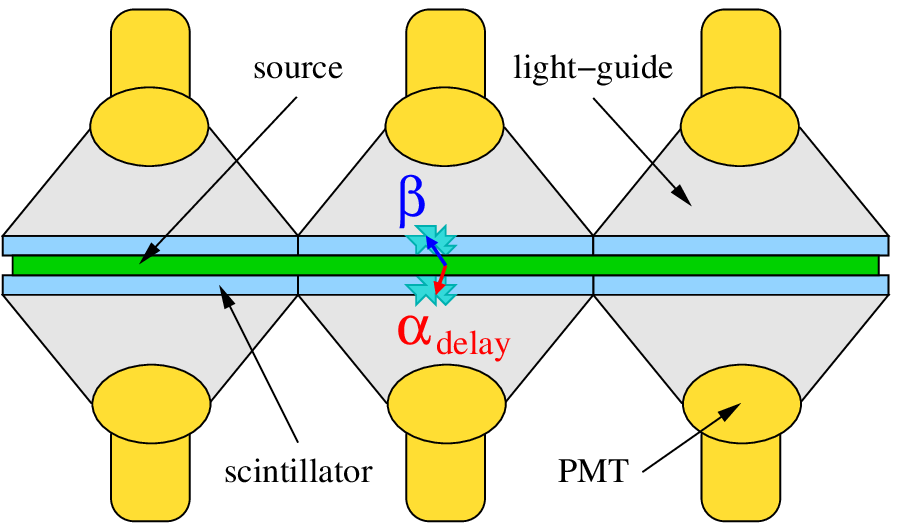}
  \hspace{1.5cm}
  \includegraphics[scale=0.45]{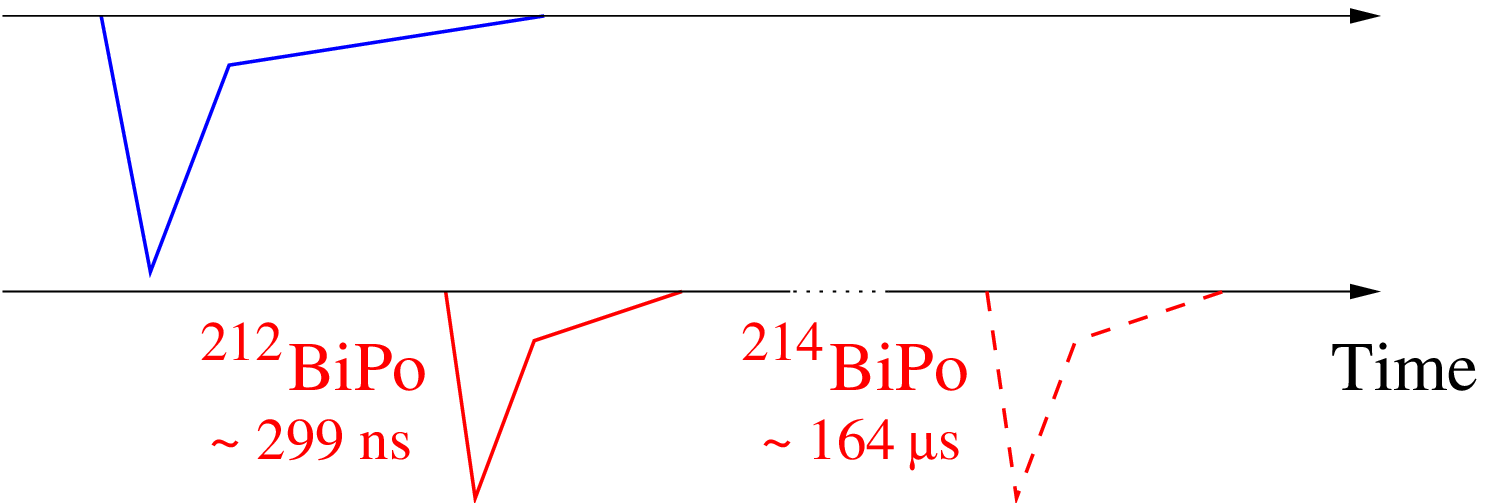}
  \caption{The BiPo detection principle with plastic scintillators and the time signals seen with PMTs for a {\it back-to-back} BiPo event.}
  \label{fig:bipo-event}
\end{figure}

The BiPo experimental intent is to install the double beta decay source foil of interest between two thin ultra-radiopure organic plastic scintillators. The $^{212}$Bi ($^{208}$Tl) and $^{214}$Bi contaminations inside the foil are then measured by detecting the $\beta$ decay as an energy deposition in one scintillator without a coincidence from the opposite side, and the delayed $\alpha$ as a delayed signal in the second opposite scintillator without a coincidence in the first one. Such a BiPo event is identified as a {\it back-to-back} event since the $\beta$ and $\alpha$ enter different scintillators on opposite sides of the foil. The timing of the delayed $\alpha$ depends on the isotope to be measured (Figure~\ref{fig:bipo-event}).
The energy of the delayed $\alpha$ provides information on whether the contamination is on the surface or in the bulk of the foil.

A second topology of BiPo events can in principle be used. This involves the {\it same-side} BiPo events for which the prompt $\beta$ signal and the delayed $\alpha$ signal are detected in the same scintillator without a coincidence signal in the scintillator on of the opposite side. The detection of the {\it same-side} events would increase by about 50\% the BiPo efficiency. However, it has been shown in \cite{bipo1} that the level of background measured in BiPo-1 using the {\it same-side} topology is much larger than the one measured in the {\it back-to-back} topology. For this reason, only {\it back-to-back} topology will be used for the rest of the BiPo-1 analysis presented in this paper.

\section{The different components of the background}
\label{sec:backgrounds}

\subsection{Random coincidences}
\label{subsec:randomcoincidences}

The first limitation of the BiPo detector is the rate of random coincidences between the two scintillators giving a signal within the delay time window (Figure~\ref{fig:bipo-background}a).
A delay time window equal to about three times the half-life of the  $\alpha$ decay is choosen (1~$\mu$s for $^{212}$Bi and 500~$\mu$s for $^{214}$Bi) in order to contain a large part of the signal and to minimize the random coincidence.
The single counting rate is dominated by Compton electrons due to external $\gamma$. BiPo-1 were then built with thin scintillators, with low radioactivity materials and installed inside a low radioactivity shield in an underground laboratory in order to reduce the $\gamma$ event rate. Additionally pulse shape analysis of the delayed signal is performed in order to discriminate between  electrons and  $\alpha$ events, thus rejecting random Compton electron coincidences due to external $\gamma$. 

\subsection{Radiopurity of the scintillators}
\label{subsec:scintradiopurity}

\begin{figure}[htb]
  \centering
  \includegraphics[scale=0.65]{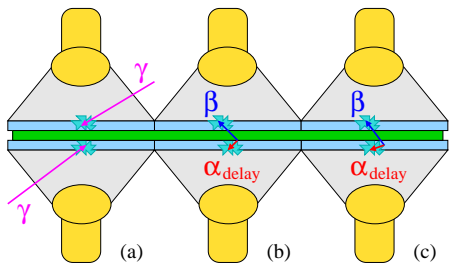}
  \caption{Illustration of the possible sources of background: (a) random coincidences due to the $\gamma$ flux, (b) $^{212}$Bi or $^{214}$Bi contamination on the surface of the scintillators and (c) $^{212}$Bi or $^{214}$Bi  contamination in the volume of the scintillator.}
  \label{fig:bipo-background}
\end{figure}

The second source of background that mimics a BiPo event comes from $^{212}$Bi or $^{214}$Bi contamination on the surface of the scintillator that is in contact with the foil. This surface contamination of the scintillators produces a signal indistinguishable from the true BiPo signal coming from the source foil, as shown in Figure~\ref{fig:bipo-background}b. 

In principle, $^{212}$Bi or $^{214}$Bi contamination in the scintillator volume (bulk contamination) is not a source of background, because the emitted electron should trigger one scintillator block before escaping and entering the second one, as shown in Figure~\ref{fig:bipo-background}c. The two fired scintillator blocks are in coincidence and this background event is rejected.
However, if the contamination is not deep enough inside the scintillator but quite near the surface, the electron from the $^{212}$Bi-$\beta$ decay will escape the first scintillator and will fire the second one without depositing enough energy to trigger the first one. It will appear exactly like a BiPo event emitted from the foil.

\subsection{Radon and Thoron}
\label{subsec:radonthoron}

Other possible sources of background are via thoron ($^{220}$Rn) or radon ($^{222}$Rn) contamination of the gas between the foil and the scintillators. Thoron and radon decay to $^{212}$Bi and $^{214}$Bi respectively and a bismuth contamination on the surface of the scintillators is thus observed.
In order to suppress this source of background, radiopure gas and materials without thoron and radon degazing around the scintillators are required.

\section{The BiPo-1 prototype}
\label{sec:bipo1prototype}

\begin{figure}[htb]
  \centering
  \includegraphics[scale=0.36]{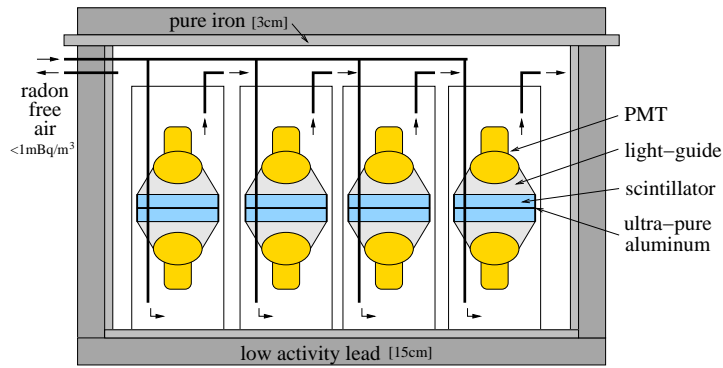}
  \caption{Schematic view of the BiPo-1 prototype inside its shield.}
  \label{fig:bipo1-setup}
\end{figure}

The BiPo-1 detector is composed of 20 similar modules. Each module consists of a gas and light tight box, containing two thin polystyrene-based scintillator plates of dimension 200$\times$200$\times$3~mm$^3$ which are placed face-to-face. Each scintillator plate is coupled to a low radioactivity 5'' photomultiplier (PMT R6594-MOD from Hamamatsu) by a UV Polymethyl Methacrylate light guide (Figure~\ref{fig:bipo1-setup}). Each module has a sensitive surface area of 0.04~m$^2$. The scintillators have been prepared with a mono-diamond tool from scintillator blocks produced by JINR (Dubna, Russia) for NEMO-3. The surface of the scintillators facing the source foil has been covered with 200~nm of evaporated ultra-pure aluminum in order to optically isolate each scintillator and to improve the light collection efficiency. The entrance surface of the scintillators has been carefully cleaned before and after aluminium deposition using the  following cleaning sequence: acetic acid, ultra pure water bath, di-propanol and finally a second ultra pure water bath. The sides of the scintillators and light guides are covered with a 0.2~mm thick Teflon layer for light diffusion. Materials of the detector have been selected by HPGe measurements to confirm their high radiopurity. The modules are shielded by 15~cm of low activity lead which addresses the external $\gamma$ flux. The upper part of the shield which supports the lead is a pure iron plate 3~cm thick. Radon-free air ($\mathcal{A}$(radon)~$<$~1~mBq/m$^3$) flushes the volume of each module and also the inner volume of the shield. 

The first three modules of BiPo-1 were initially installed in the new Canfranc Underground Laboratory in Spain. Due to the temporary closure of this Laboratory, BiPo-1 was completed in the Modane Underground Laboratory where it has been running since May 2008.

Photomultiplier signals are sampled with MATACQ VME digitizer boards~\cite{breton}, during a 2.5~$\mu$s window with a high sampling rate (1~GS/s), 12-bit resolution and a dynamic range (1~volt). The level of electronic noise is $\sigma$=250~$\mu$V. The single photoelectron level is 1~mV.
The acquisition is triggered each time a PMT pulse reaches 50~mV (corresponding to 100~keV). 

More informations about the BiPo-1 prototype, the radiopurity of its components, the calibrations and the detection efficiency can be found in \cite{bipo1}.

\section{Results from the BiPo-1 detector}
\label{sec:bipo1results}

\subsection{Experimental verification of the BiPo-1 technique with a calibrated aluminium foil}
\label{subsec:principlevalidation}

The first BiPo-1 module was dedicated to testing  the detection of bulk $^{212}$Bi contamination in a calibrated foil. A 150~$\mu$m thick aluminium foil (40~mg/cm$^2$) with an activity $\mathcal{A}$($^{212}$Bi$\rightarrow ^{212}$Po)~=~0.19~$\pm$~0.04~Bq/kg was first measured with low background HPGe detectors, and then installed between the two scintillators of the BiPo-1 module.

\begin{figure}[htb]
  \centering
  \includegraphics[scale=0.21, angle=270]{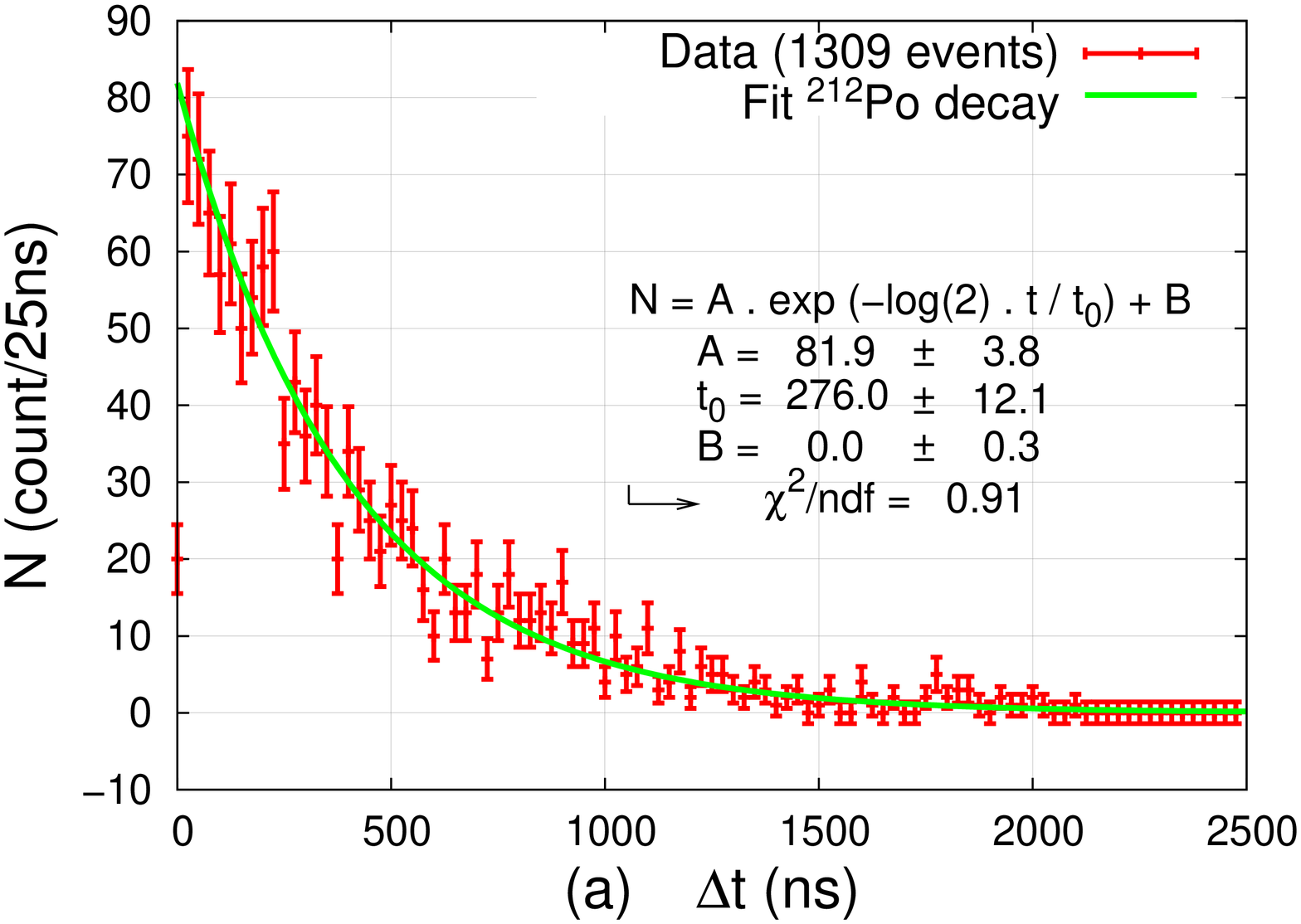}
  \includegraphics[scale=0.21, angle=270]{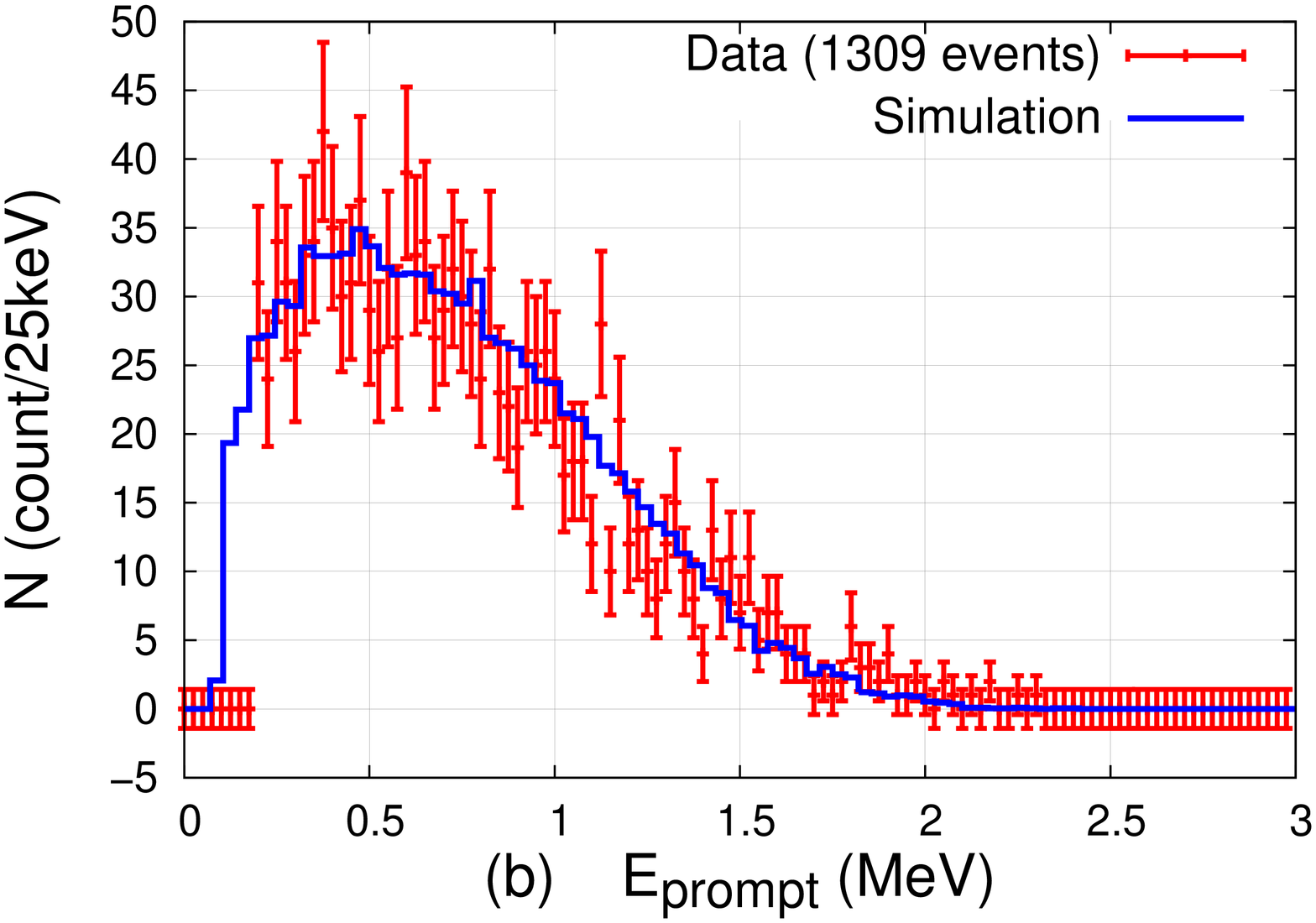}
  \includegraphics[scale=0.21, angle=270]{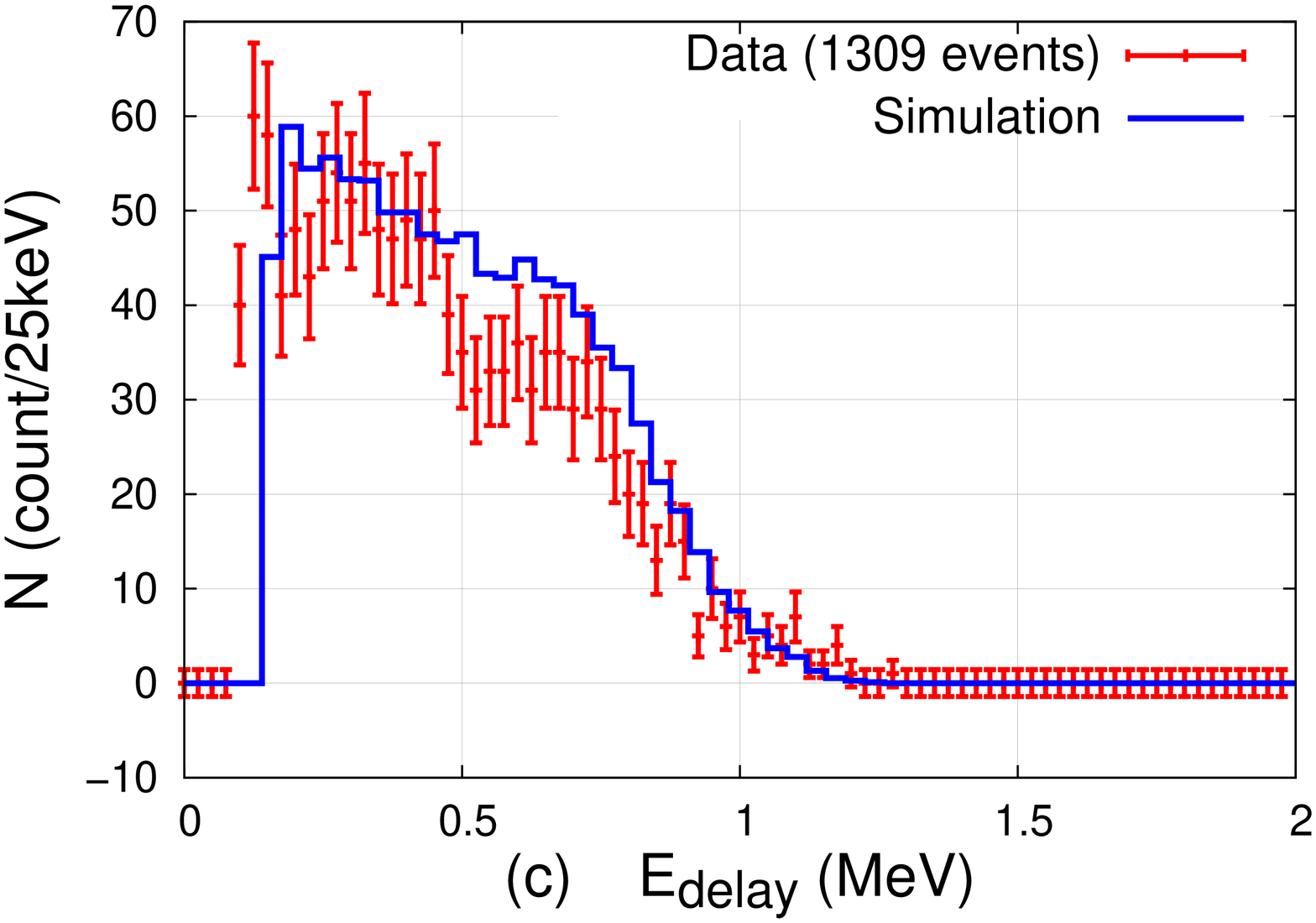}
  \caption{(a) The delay time distribution between the $\beta$ and $\alpha$ decays, (b) energy distribution of the prompt $\beta$ and (c) energy distribution of the delayed $\alpha$, for the 1309 BiPo events detected with the calibrated aluminium foil.}
  \label{fig:bipo1-c1-results}
\end{figure}

After 160~days of data collection, a total of 1309 {\it back-to-back} BiPo events were detected. Taking into account the 3.4$\%$ efficiency calculated by GEANT4 simulations with a 20\% systematic error \cite{bipo1}, this corresponds to an activity of $\mathcal{A}$($^{212}$Bi$\rightarrow ^{212}$Po)~$ = 0.16 \pm 0.005 (stat) \pm 0.03(syst)$~Bq/kg, in good agreement with the HPGe measurement. The distribution of the delay time between the two signals is presented in Figure~\ref{fig:bipo1-c1-results}. The half-life obtained from the fit is T$_{1/2}=276 \pm 12 (stat)$~ns. This measured half-life is in agreement with the experimental weighted average value of 299~ns for $^{212}$Po~\cite{po212}. These results confirm the measurement principle and the calculated efficiency. 

The energy spectra of the prompt $\beta$ and the delayed $\alpha$ are presented in Figure~\ref{fig:bipo1-c1-results}. There is good agreement with the expected spectra calculated by simulations. The  energy spectrum of the first  signal corresponds to a typical $\beta$ spectrum with $Q_{\beta}=2.25$~MeV. The energy spectrum of the delayed signal goes up to 1~MeV as expected for the $\alpha$ of 8.78~MeV from $^{212}$Po and a  quenching factor of Q$_f\sim$~9 at 8.78 MeV \cite{bipo1}. No evidence of a peak at the 1~MeV endpoint in the energy spectrum of the delayed $\alpha$ indicates that the radioactive contamination is inside the volume and not on the surface of the aluminium foil.

\subsection{Discrimination of $\beta$ and $\alpha$ particles}
\label{subsec:eadiscrimination}

The excitation of the scintillator depends on many factors including the energy loss density and a larger $dE/dx$ for alphas enhances the slow component of the decay curve. This behavior has been already observed in organic liquid scintillator~\cite{borexino}. We show here that this difference is also observed in polystyrene-based plastic scintillator and can be used to discriminate alphas to electrons.

\begin{figure}[htb]
  \centering
  \includegraphics[scale=0.34]{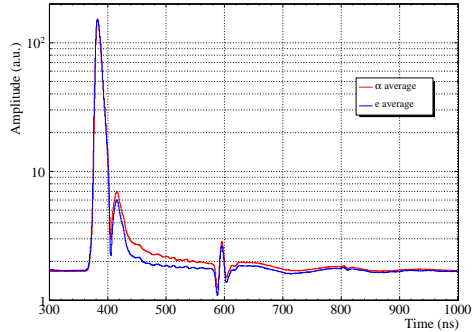}
  \caption{Average PMT signals obtained with a BiPo-1 module, in dashed line for 1 MeV electrons ($^{207}$Bi source), and in grey line for 5.5 MeV $\alpha$  ($^{241}$Am). The amplitudes have been normalized to the first peak. The secondary peaks are due to electronic bounces in the PMT HV divider due to an imperfect impedance matching.}
  \label{fig:bipo1-discri-signal}
\end{figure}

The average PMT signal obtained with a BiPo-1 module for electrons ($^{207}$Bi source) or $\alpha$  ($^{241}$Am source) is presented in Figure~\ref{fig:bipo1-discri-signal}. A small but significantly larger component in the tail of the signal is well observed for $\alpha$ particles compared to electrons. A pulse shape discrimination was developed using this set of data. The discrimination factor $\chi$ is defined as the ratio of the charge $q$ in the slow component to the total charge $Q$ of the signal. The charge $q$ is integrated from 15~ns after the signal peak to 900~ns. This integration window was optimized in order to maximize the discrimination. 

\begin{figure}[htb]
  \centering
  \includegraphics[scale=0.28]{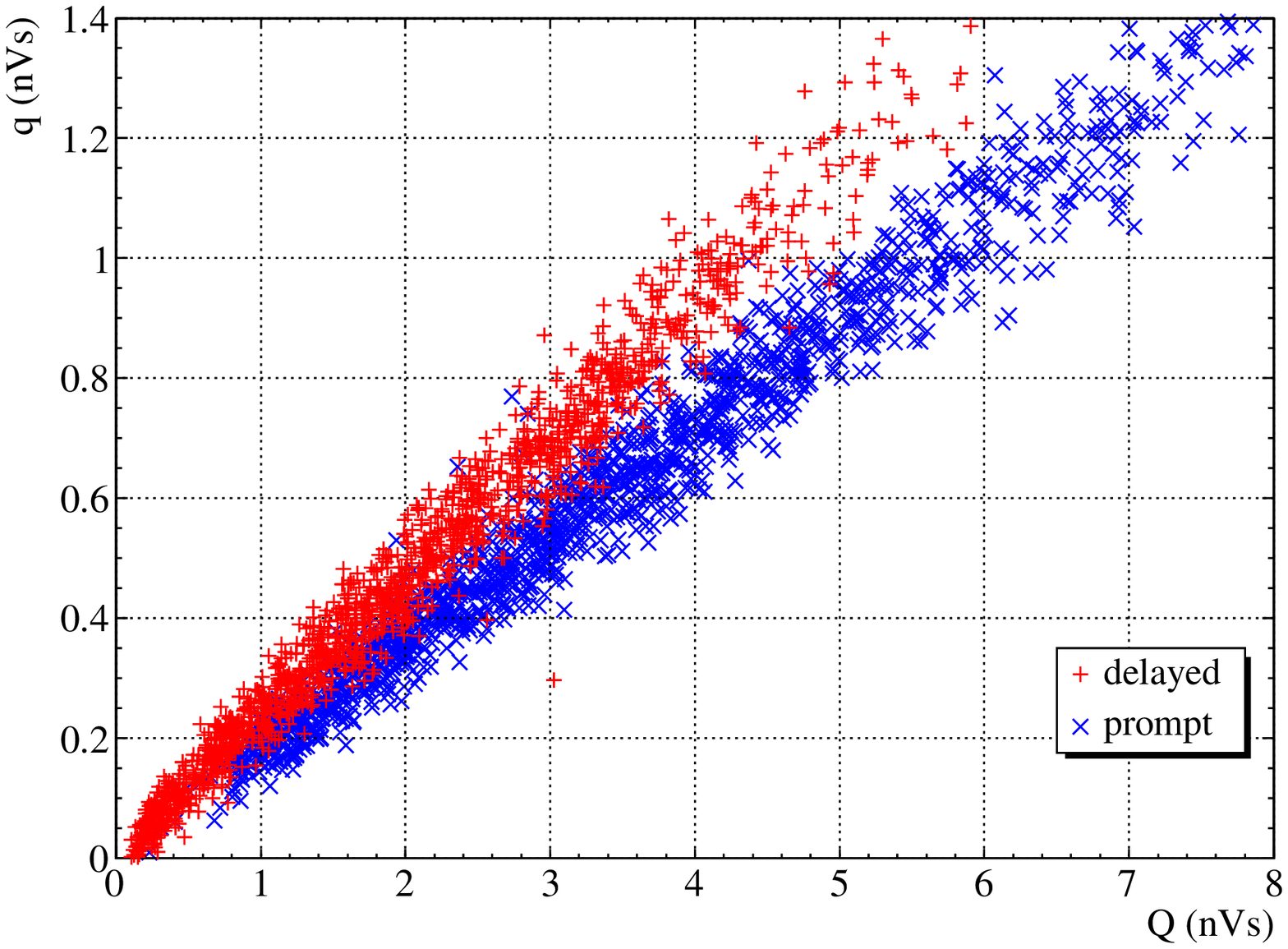}
  \includegraphics[scale=0.28]{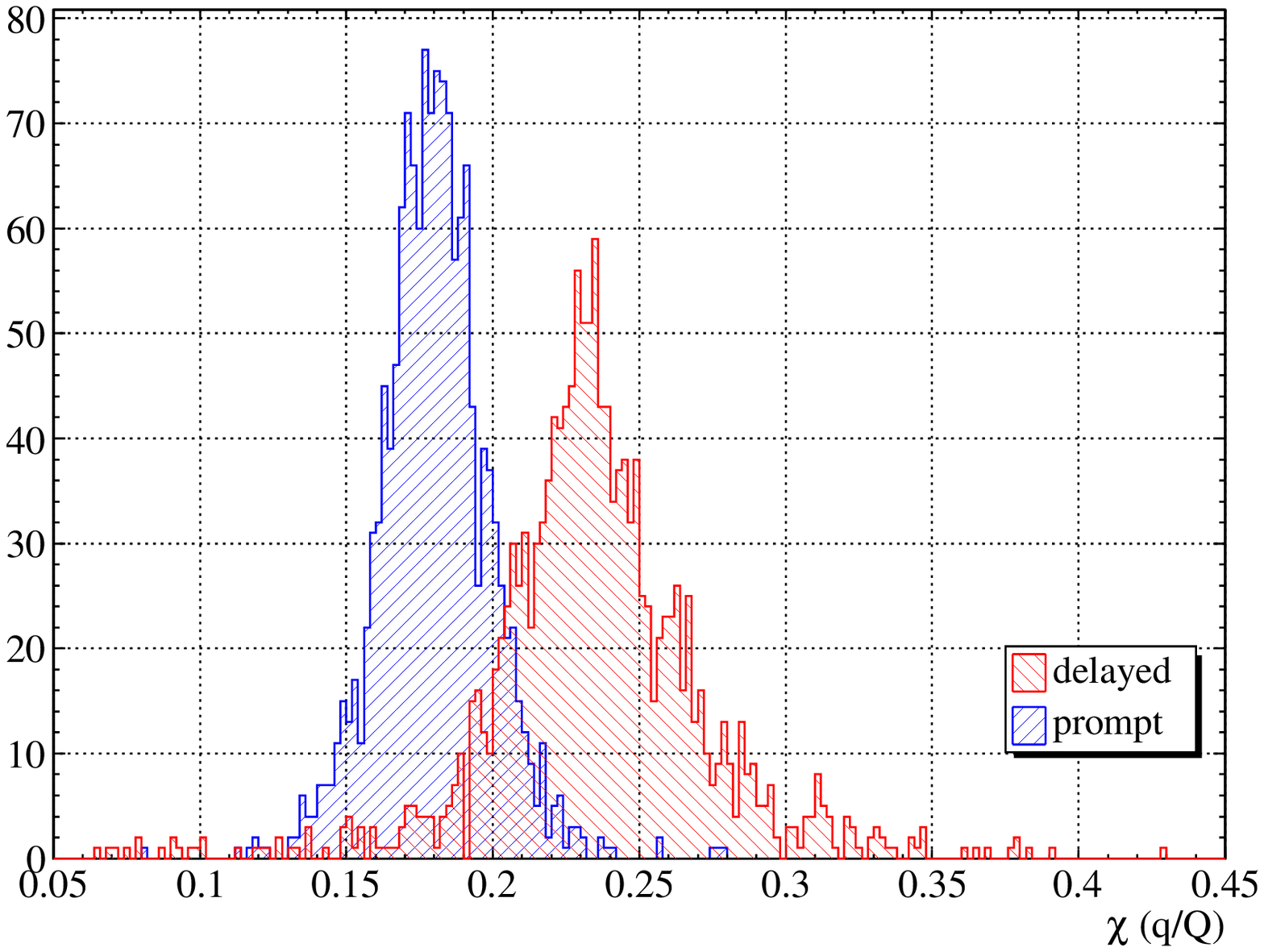}
  \includegraphics[scale=0.28]{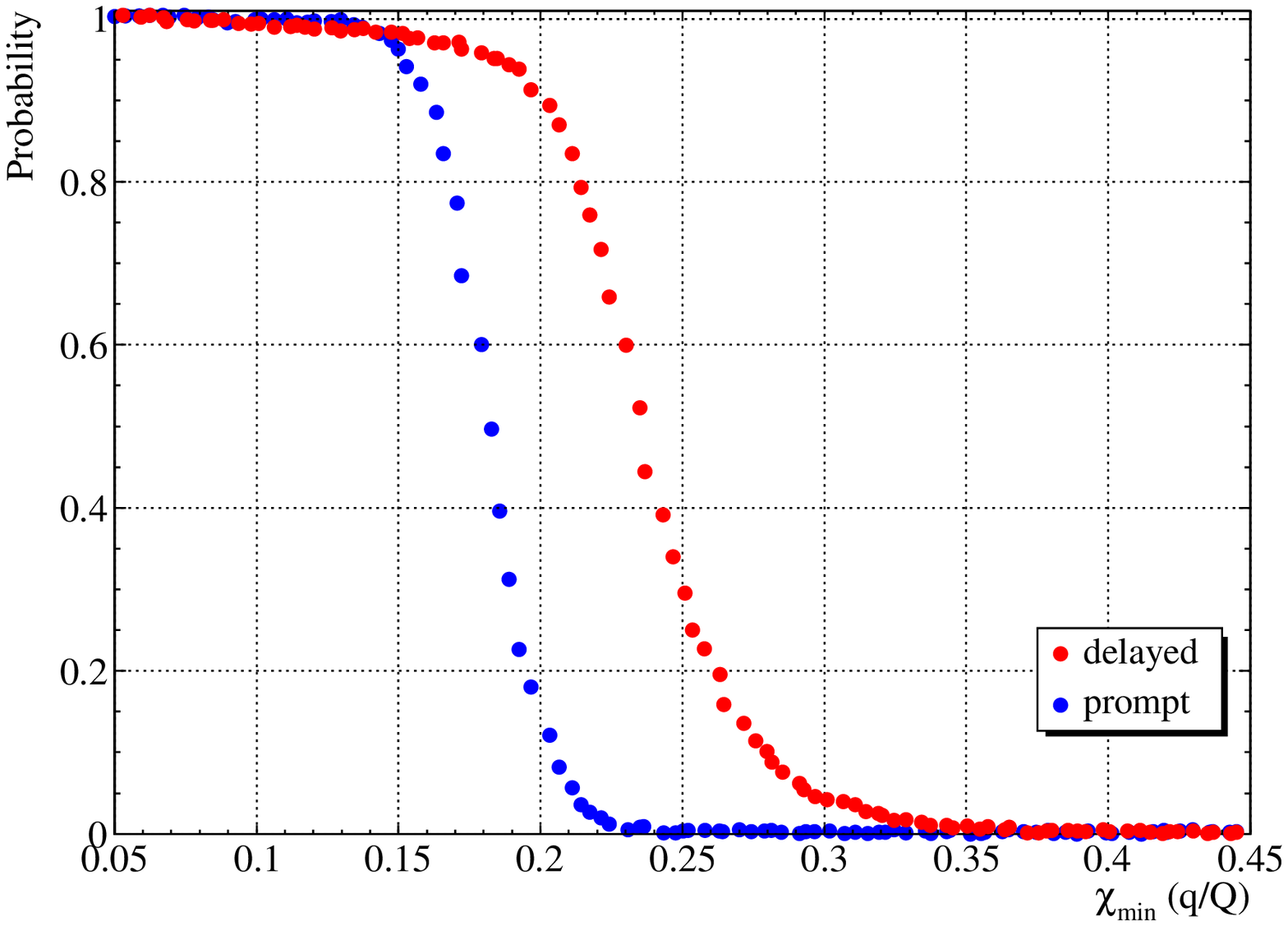}
  \caption{Discrimination of the prompt $\beta$  and delayed $\alpha$ signals using the BiPo events measured with the calibrated aluminium foil: (Left) Distribution of the charge $q$ in the slow component of the signal as a function of the total charge $Q$ of the signal; (Center) Distribution of the discrimination factor $\chi = {q \over Q}$; (Right) Probability of selecting an electron (black dot) or an $\alpha$ (grey triangle) as a function of the lower limit $\chi_{min}$ set by the discrimination factor $\chi$. The criterion $\chi > 0.2$ allows one to keep 90\% of the $\alpha$ particles and to reject 85\% of the electrons.}
  \label{fig:bipo1-discri-caps1}
\end{figure}

This electron and $\alpha$ pulse shape discrimination has been applied to the set of the BiPo events detected with the calibrated aluminium foil inside a first BiPo-1 module in order to calculate its global efficiency. As shown in Figure~\ref{fig:bipo1-discri-caps1}, a good separation is observed for  $\beta$ and $\alpha$ signals although the discrimination becomes less efficient at low energy. By selecting a discrimination factor of $\chi > 0.2$ as applied to the delayed signal allows one to keep 90\% of the true BiPo events and reject 85\% of the random coincidences as is shown in Figure \ref{fig:bipo1-discri-caps1}. 

\subsection {Random coincidences}
\label{sec:random-coinc}

The single counting rate $\tau_0$ of a BiPo-1 scintillator without any coincidence in the opposite scintillator of the same module, was measured in the Modane Underground Laboratory with an energy threshold of 150 keV and $\tau_0 \approx 20$~mHz. This counting rate which is dominated by Compton electrons produced by external $\gamma$'s is stable and independent of the modules. It corresponds to an expected number of coincidences for the $^{212}$Bi measurement of $N_{coinc}$($^{212}$BiPo)~$= 2.10^{-3}$ events/(BiPo-1 module)/month, a negligible background. 

This result, extrapolated to the larger final BiPo-3 detector is low enough to reach a sensitivity of 2~$\mu$Bq/kg in $^{208}$Tl as required by the SuperNEMO experiment. 

However, for the $^{214}$Bi measurement, the number of random coincidences $N_{coinc}$($^{214}$BiPo) becomes too large  because of the longer half-life of $^{214}$Po. This background can be reduced by applying the electron-$\alpha$ discrimination defined earlier to the delayed signal in order to select only delayed $\alpha$ particles and reject the Compton electron coincidences. The number of random coincidences after electrons-$\alpha$ discrimination becomes acceptable (reduction by a factor 7). Given no other background component, this would correspond to a sensitivity of about 10~$\mu$Bq in $^{214}$Bi as required by the SuperNEMO experiment. 

\subsection {Measurement of the scintillator bulk radiopurity}
\label{sec:bulk-bkg}

A BiPo-1 module has been dedicated to measure the bulk radiopurity of the organic polystyrene-based plastic scintillators produced by the JINR (Dubna, Russia) and used in BiPo-1. This module is similar to a standard BiPo-1 module except that it is equipped with two thicker scintillator blocks 20$\times$20$\times$10~cm$^3$ each, that are wrapped with aluminized Mylar. One of these blocks is from the NEMO-3 batch production used in BiPo-1. The second one is from a newer manufacturing process at JINR in 2007. 

The $^{212}$Bi contamination inside the scintillator blocks is recognized as a prompt signal from one PMT and a delayed signal of up to 1~$\mu$s from the same PMT. 
Since the delayed $\alpha$ is fully contained in the scintillator, its deposited energy in scintillation is expected to be around 1 MeV, due to the quenching factor. Thus, it is required that the energy of the delayed signal be greater than 700 keV (5 sigma less than 1 MeV). 
After 141 days of data collection, 10 events have been detected in the scintillator block from the NEMO-3 batch production and 24 events in the block from the new production. These numbers are in agreement with the expected number of BiPo events emitted from the aluminized Mylar surrounding the scintillator. However, if it is assumed that all the detected BiPo events come from a bulk contamination of the scintillators, conservative upper limits on the contaminations from $^{208}$Tl in the scintillators are:
\begin{itemize}
\item NEMO-3 batch production used in BiPo-1: $\mathcal{A}$($^{208}$Tl)~$\leq$~0.13~$\mu$Bq/kg
\item New JINR production: $\mathcal{A}$($^{208}$Tl)~$\leq$~0.3~$\mu$Bq/kg
\end{itemize}

Given a bulk contamination in the scintillator of $\mathcal{A}$($^{208}$Tl)~$=$~0.13~$\mu$Bq/kg, the expected background level for BiPo-1, calculated by Monte Carlo simulations, is equal to 0.003 {\it back-to-back} BiPo events/month/module. For a larger BiPo-3 detector with a sensitive surface of 3.25~m$^2$, this background is still low enough to reach the radiopurity performance required by the SuperNEMO experiment. 

\subsection{Measurement of the scintillators surface radiopurity}
\label{sec:surf-bkg}

Thirteen BiPo-1 modules have been used for background measurements of the scintillator surfaces. The scintillators were placed face-to-face without a foil between them. After 488 days of data collection, a total of 42 {\it back-to-back} BiPo events have been observed. One module appeared to be more polluted with 12 events detected in this  module. The other events were uniformly distributed in the other 12 modules and in arrival time. The contaminated module has been removed from the analysis. For the 12 remaining modules, corresponding to an integrated scintillator surface of 468~m$^2$.days, only 30 {\it back-to-back} BiPo events have been observed. For the same period of 488 days of data collection, the expected number of random coincidences is about 0.2  (see section~\ref{sec:random-coinc}) and the expected background due to a bulk contamination in the scintillators has a maximum of 0.6 BiPo events (see section~\ref{sec:bulk-bkg}). Thus, the background observed in BiPo-1 corresponds to a $^{212}$Bi bismuth contamination on the surface of the scintillators. Taking into account the 27\% efficiency, calculated by simulations, to detect the BiPo cascade from a $^{212}$Bi pollution on the surface of the scintillators with a 20\% systematic error, this corresponds to a surface background of the BiPo-1 prototype of $\mathcal{A}$($^{208}$Tl)~$= 1.5 \pm 0.3 (stat) \pm 0.3 (syst)$~$\mu$Bq/m$^2$ in $^{208}$Tl.
 
\begin{figure}[htb]
  \centering
  \includegraphics[scale=0.22, angle=270]{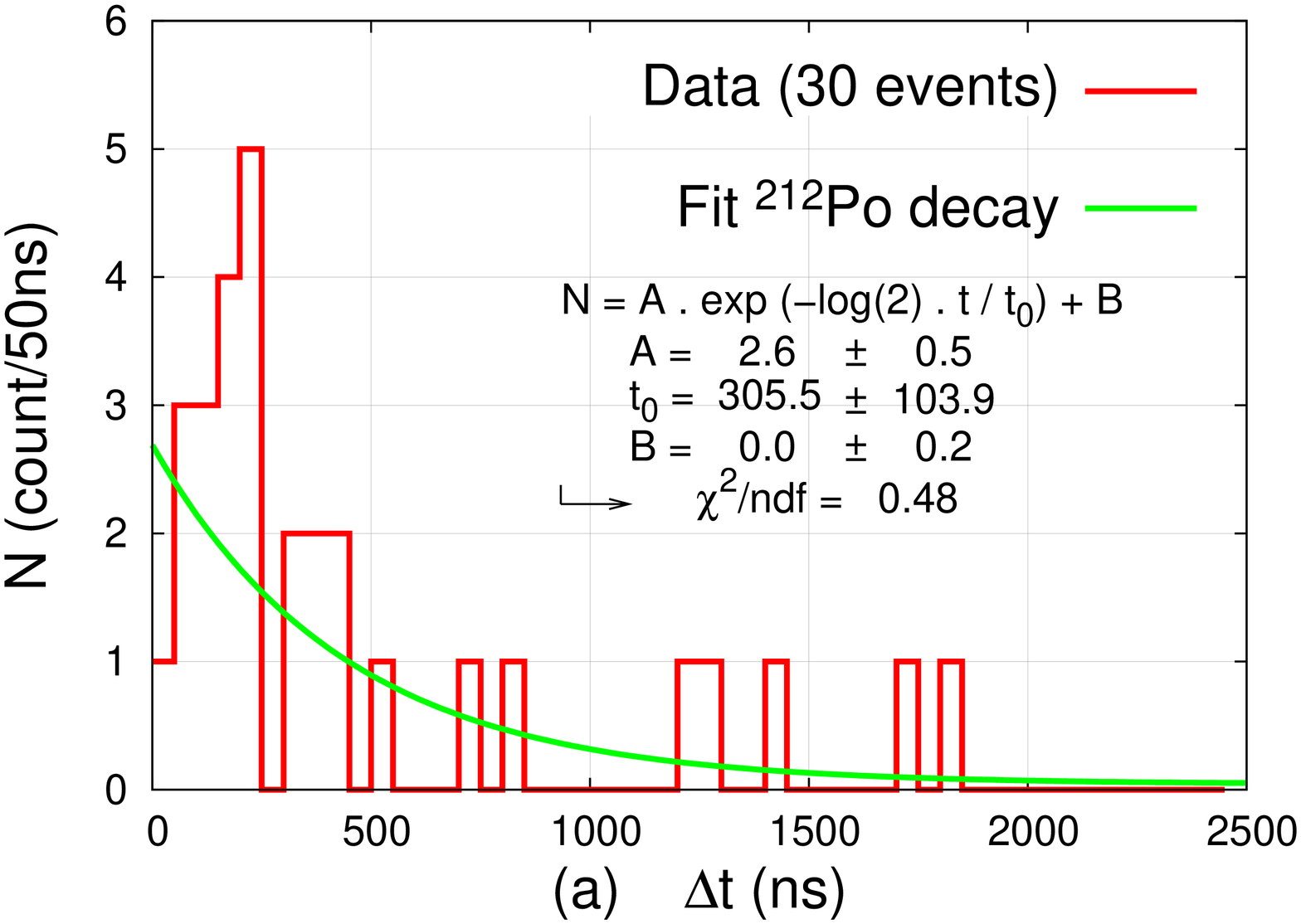}
  \includegraphics[scale=0.22, angle=270]{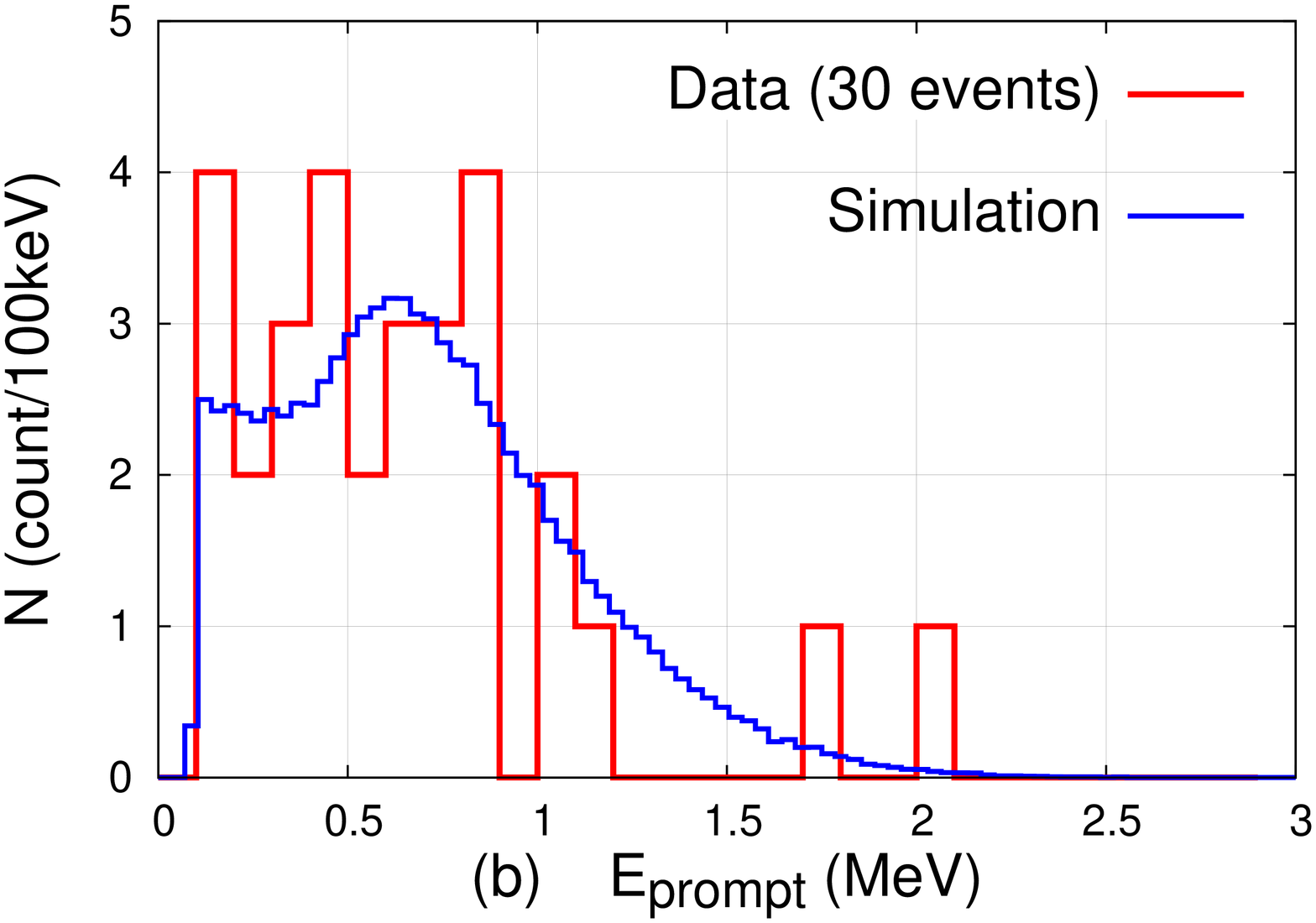}
  \includegraphics[scale=0.22, angle=270]{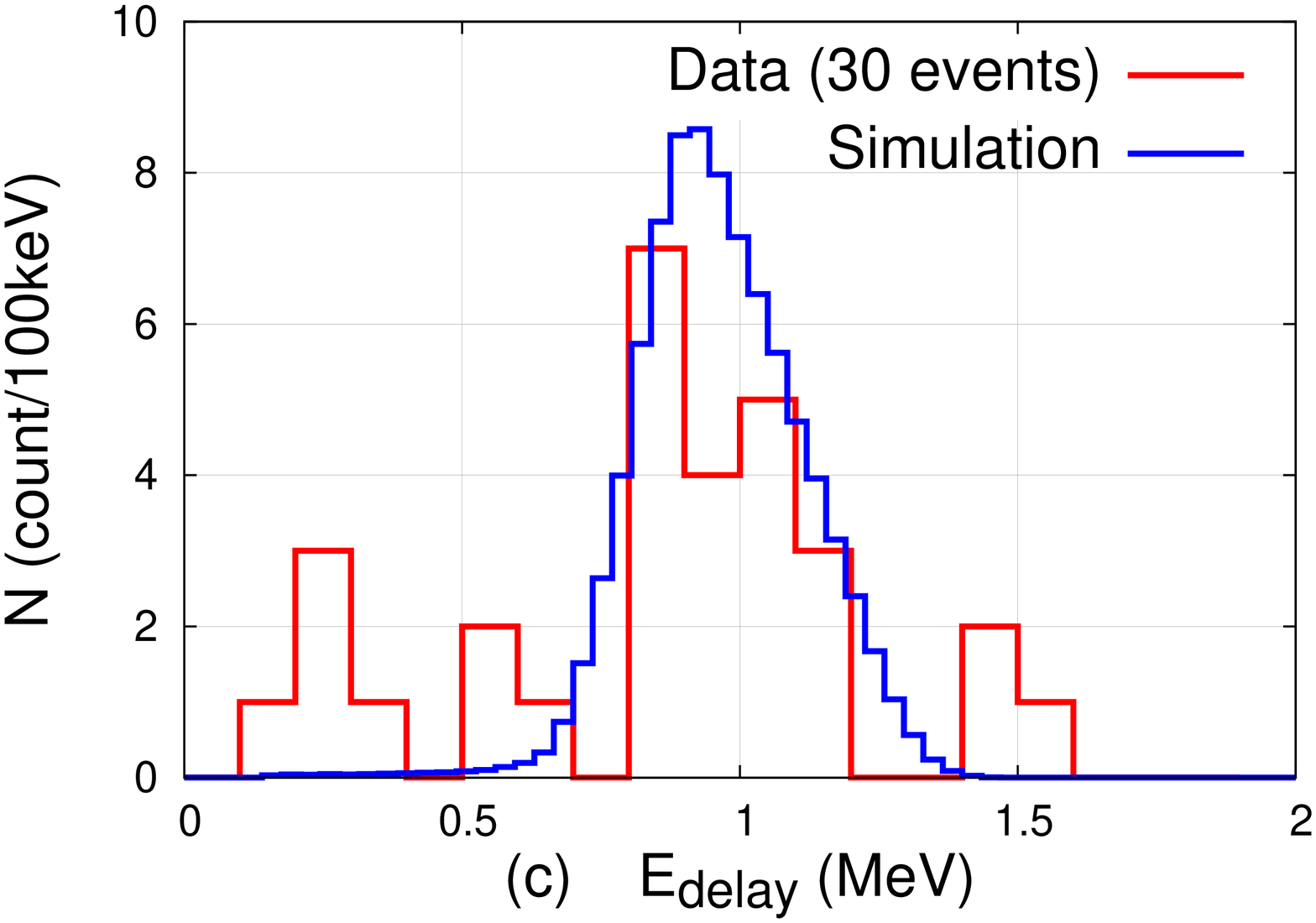}
  \caption{Distributions (a) of the delay time, (b) of the energy of the prompt signal and (c) the energy of the delayed signal, shown in grey for the 30 BiPo background events observed during 423 days of data collected with 12 BiPo-1 modules, and in black for the Monte-Carlo simulations. The black curve in the delay time distribution is the results of the exponential decay fit.}
  \label{fig:bipo-surfbkg} 
\end{figure}

The distribution of the delay time between the two signals is presented in Figure~\ref{fig:bipo-surfbkg}. Despite the low statistics, it is compatible with an exponential decay distribution with a half-life of T$_{1/2}=305 \pm 104 (stat)$~ns. The energy spectra of the prompt $\beta$ and delayed $\alpha$ signals, also presented in Figure~\ref{fig:bipo-surfbkg}, are in good agreement with the expected spectra calculated by the simulations. The energy distribution of the delayed signal is centered around 1 MeV as expected for the 8.78~MeV $\alpha$ emitted from $^{212}$Po on the surface of the scintillators with a quenching factor of 9. The origin of the few events at lower energies is still unknown.

There are several possible origins of the surface background observed with BiPo-1. It might be due to contaminations either during the surface machining of the scintillator plates, during the cleaning procedure of the scintillators or during the evaporation of the aluminium on the entrance face of the scintillators\footnote{The evaporation was done with a sputtering setup which was carefully cleaned.}. A possible contamination of the deposited aluminium is unlikely to be the origin of the background. The radiopurity of this aluminium has been measured with low background HPGe and $\mathcal{A}$($^{208}$Tl)~$<$~0.2~mBq/kg. This radiopurity corresponds to less than 2 BiPo events, which is much less than the 30 background events observed in BiPo-1. Another possible origin of the observed background is a thoron contamination between the two scintillators. Considering a typical gap of air of 200~$\mu$m, the 30 observed background events would correspond to a thoron activity of the air of $\mathcal{A}$(thoron)~$\sim$~30~mBq/m$^3$. However, this level of contamination seems too large compared to an estimation of thoron emanation from the PMTs. 

\section{The BiPo-3 Detector}
\label{sec:bipo3}

\begin{figure}[htb]
  \centering
  \includegraphics[scale=0.36]{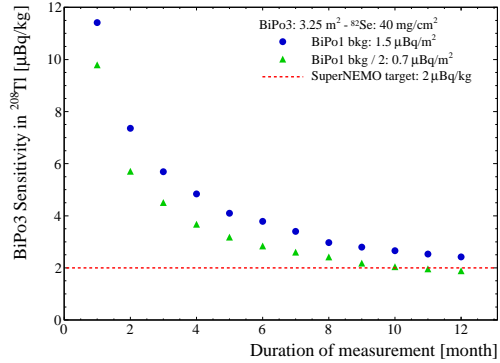}
  \caption{Sensitivity in $^{208}$Tl of the BiPo-3 detector as a function of the time of measurement with the BiPo-1 scintillator surface background or with an expected reduced background. A sensitivity of $\mathcal{A}$($^{208}$Tl)~$<$~3-4~$\mu$Bq/kg (90\% C.L.) can be reached with a six month measurement.}
  \label{fig:bipo3-sens} 
\end{figure}

Given the good results of the BiPo-1 prototype and the necessity to have a larger measurement surface for the development of the SuperNEMO double beta sources, a medium-size BiPo-3 detector, using the BiPo-1 technique, with a total surface area of 3.25m$^2$ is under construction in LAL (Orsay). The size of this detector is a compromise to fulfill the requirements of a relatively small detector (and thus a relatively small number of readout channels) and a good enough detector sensitivity in order to quantify the background in the first SuperNEMO selenium double beta foil sources used for the SuperNEMO demonstrator. Given the BiPo-1 background, it is shown in Figure~\ref{fig:bipo3-sens} that the BiPo-3 expected sensitivity in $^{208}$Tl will be $\mathcal{A}$($^{208}$Tl)~$<$~3-4~$\mu$Bq/kg (90\% C.L.) with a six month measurement.

\subsection{Description of the BiPo-3 Detector}
\label{sec:bipo3-detector}

\begin{figure}[htb]
  \centering
  \includegraphics[height=4.0cm]{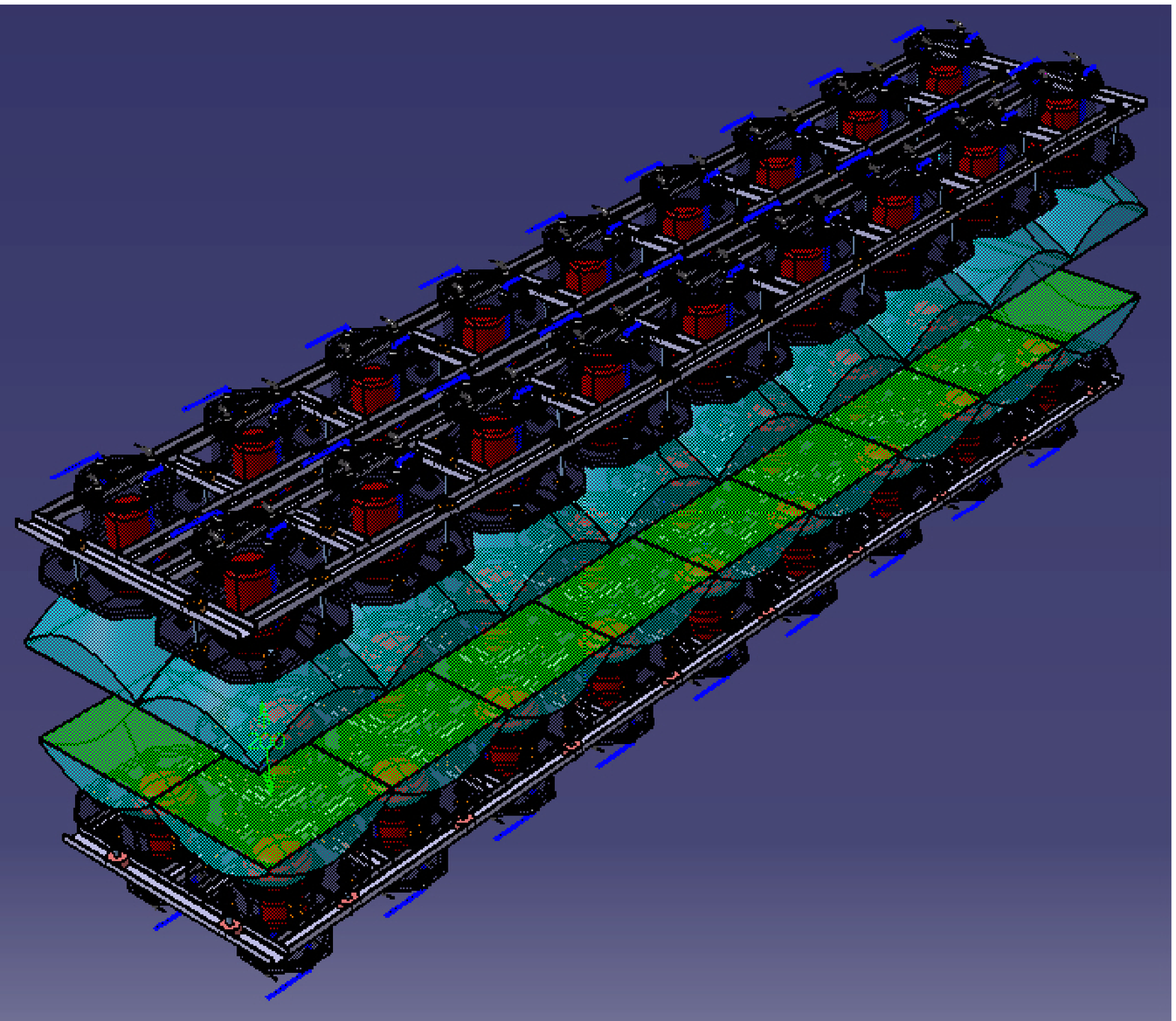}\hspace{5mm}
  \includegraphics[height=4.0cm]{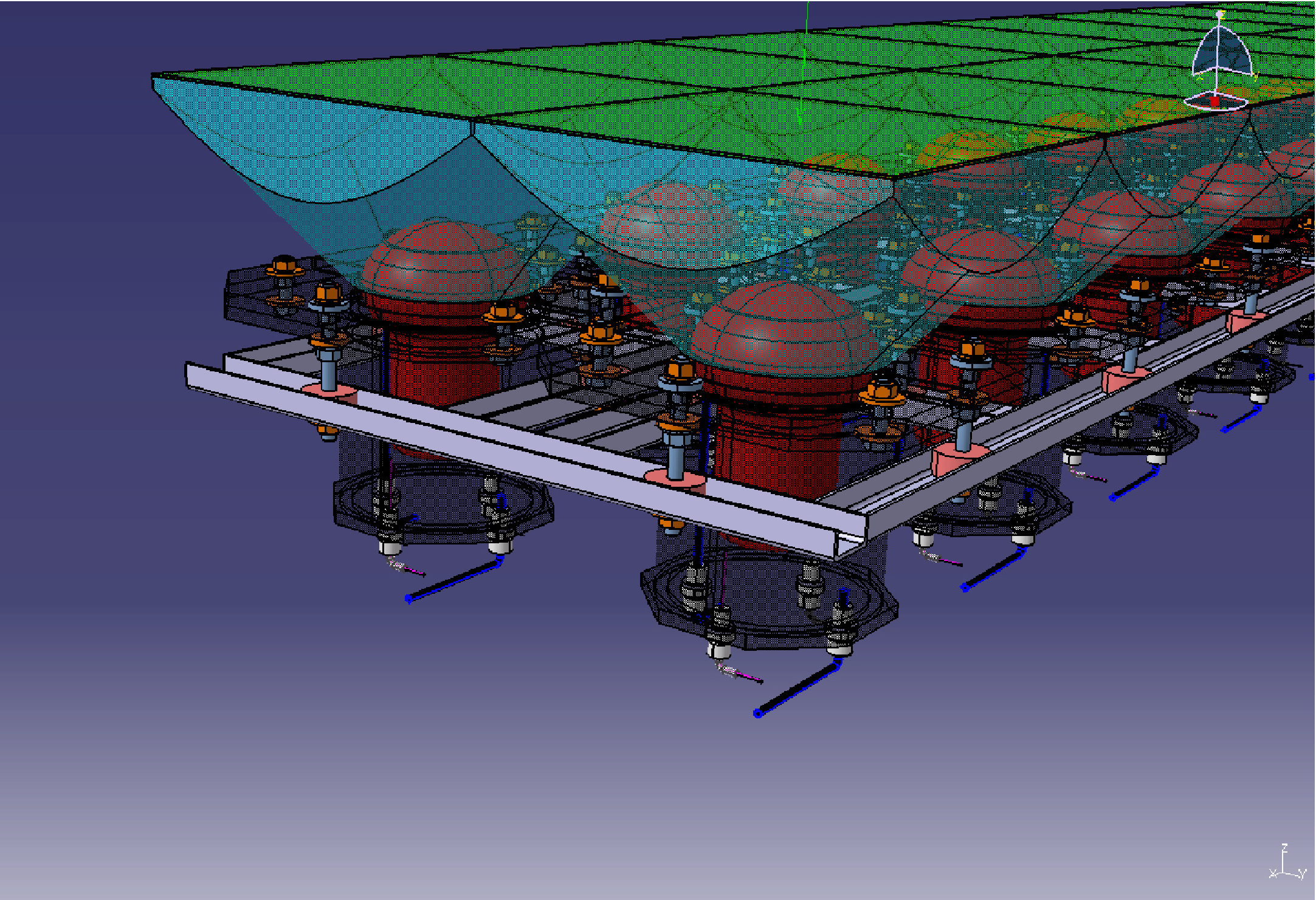}\hspace{5mm}
  \includegraphics[height=4.0cm]{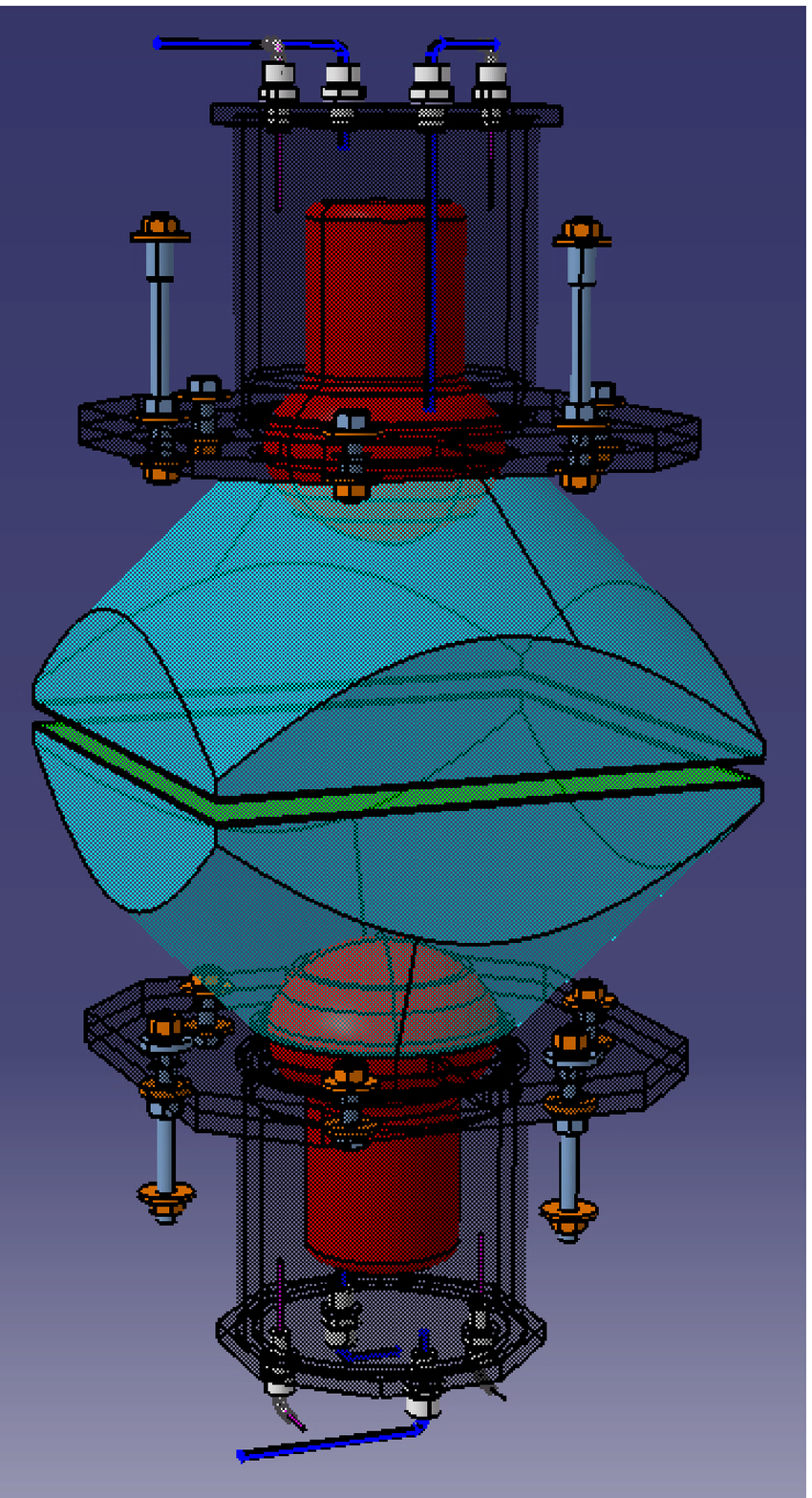}
  \caption{Drawings of one module of the BiPo-3 prototype showing the optical modules (scintillator - PMT -light guide) facing each other.}
  \label{fig:bipo3-detector} 
\end{figure}

The BiPo-3 detector is composed of two modules. Each module consists of two rows
of aluminized polystyrene-based scintillator plates arranged face-to-face (like in BiPo-1) coupled to 5'' low radioactive Hamamatsu PMTs, with a PMMA optical guide wrapped with Tyvek (Figure~\ref{fig:bipo3-detector}). The size of each scintillator is 300x300x2 mm$^3$. This corresponds to a total number of 72 PMTs and a total detector surface of 3.25m$^2$. Each module is contained inside a polyethylene tight box and inserted inside the low radioactive shield.

Some improvements have been made compared to BiPo-1 prototype:
\begin{itemize}
  \item surface uniformity response by the optimization of the light guide shape for a better light collection
  \item radiopurity of the scintillators surface using cleaner machining procedure and protections
  \item radiopurity of the scintillators surface using a new and dedicated setup for the aluminization of the scintillators
  \item improvement of the cleanliness of the LAL clean room for BiPo-3 optical modules tests and construction
  \item radon protection with separated areas by radon-tight films and improved flushing system to remove the emanations from the PMTs and protect the volume between the scintillators were the sourve will be measured
\end{itemize}

\subsection{The BiPo-3 Prototype}
\label{sec:bipo3-prototype}

As presented before, the construction of the BiPo-3 detector was the opportunity for improvements of the BiPo-1 technique. In order to validate these changes and the new conditions of production, we constructed a BiPo-3 prototype of one pair of optical modules facing each other. This module has been installed in the shielding of BiPo-1 in the LSM in June 2010 and as been connected to BiPo-1 DAQ.

The first observation of this prototype was a high counting rate of the optical modules (few Hz compared to 20 mHz in BiPo-1) due to an unexpected noise. The origin of this noise is due to light production inside the light guide. This light was not seen in BiPo-1 light guides because the PMT gain was lower, but it appeared when the gain was set to higher values to investigate this noise. In the BiPo-3 prototype the PMT gain had to be increased because the PMMA of the light guide was not treated for enhanced UV transparency and the light collection was too low in the first tests. A new UV enhanced PMMA has been selected for the BiPo-3 detector in order to suppress this noise and work at lower PMT gains like in BiPo-1.

Because the light produced by this noise has not the same origin of the scintillation light produced in the scintillators, the pulse shape of the PMT signals are different. It was possible to completely remove this noise by pulse shape analysis with a cut on the charge over amplitude ratio of the PMT pulses. Thanks to this analysis, it was then possible to perform the tests needed with the BiPo-3 prototype. After 107.5 days of data tacking, the preliminary results on the backgrounds for the BiPo events search are:
\begin{itemize}
  \item the single counting rate is about $\tau \sim$ 10~mHz. This is better than BiPo-1 because the distance between the PMT and the scintillator has been increased for better uniformity of light collection. The probability for natural radioactivity of PMT glass $\gamma$-rays to interact is the scintillators has been reduced with this distance.
  \item two $^{212}$BiPo events have been observed corresponding to a surface contamination in $^{208}$Tl of the new BiPo-3 scintillators between 0.7~$<~\mathcal{A}$($^{208}$Tl)~$<$~7~$\mu$Bq/m$^2$ (90 \% C.L.). Because of a smaller surface, the sensitivity of BiPo-1 has not been reached yet but this level is close to the BiPo-1 background (1.5~$\mu$Bq/m$^2$). This result concerns only one module but it shows that the background level didn't increased compared to BiPo-1.
  \item eighteen $^{214}$BiPo events have been observed corresponding to a surface contamination in $^{214}$Bi of the new BiPo-3 scintillators between 25~$<~\mathcal{A}$($^{214}$Bi)~$<$~60~$\mu$Bq/m$^2$ (90 \% C.L.). This level is still high but the radon background is included because no improvements were made on this prototype before BiPo-1 results but it will be reduced for the BiPo-3 detector. Anyway we can say that the $^{214}$Bi background has been reduced, compared to BiPo-1 were this background is still under investigation, thanks to the efforts maid for the optical modules preparation. This preliminary result is a good indication that we should be able to reach the required sensitivity of 10~$\mu$Bq/kg for the $^{214}$Bi measurement of the SuperNEMO sources.
\end{itemize}
Except the noise that was rejected by pulse shape analysis, these results are a good motivation to move forward with the construction of the BiPo-3 detector. It seems also that all the efforts added for the construction of the optical modules helped us to reduce the BiPo-3 backgrounds compared to the BiPo-1 results.

A new BiPo-3 prototype, with UV enhanced PMMA light guides is under construction and will be installed in Canfranc Underground Laboratory in the beginning of 2011. With this prototype we will confirm that the noise has been removed and we will remeasure the backgrounds to validate the improvements seen with the previous prototype running in the LSM. As the underground environment will be different, we will also use this prototype to test the shielding and the new radon free air flushing system before the construction of the full BiPo-3 detector.

\section*{Conclusion}
\label{sec:conclusion}

The SuperNEMO collaboration is developing a BiPo detector using thin plastic scintillator plates in order to measure with very high sensitivity the radiopurity in $^{208}$Tl and $^{214}$Bi of the double beta decay source foils to be used in the SuperNEMO experiment.

The BiPo-1 prototype with 0.8~m$^2$ of sensitive surface has been fully operational since May 2008 in the Modane Underground Laboratory. The detection efficiency has been experimentally verified by measuring a calibrated aluminium foil. The random coincidences have been measured and are negligible for $^{208}$Tl. For $^{214}$Bi, it has been verified that random coincidences can be reduced by applying an electron-$\alpha$ discrimination with a high enough efficiency in order to reach a sensitivity of about 10~$\mu$Bq in $^{214}$Bi as required by the SuperNEMO experiment. A first phase of data collection with BiPo-1 ran until June 2009 and was dedicated to the background measurement of $^{212}$Bi ($^{208}$Tl). The most challenging background is a contamination on the surface of the scintillators in contact with the foils. After more than one year of data collection, a surface activity of $\mathcal{A}$($^{208}$Tl)~$= 1.5 \pm 0.3 (stat) \pm 0.3 (syst)$~$\mu$Bq/m$^2$ has been measured. Given this level of background, a larger BiPo detector having 12~m$^2$ of active surface area, would be able to qualify the radiopurity of the SuperNEMO selenium double beta decay foils with the required sensitivity of $\mathcal{A}$($^{208}$Tl)~$<$~2~$\mu$Bq/kg (90\% C.L.) with a six month measurement. A second phase of data started running with new electronics in July 2009 in order to measure simultaneously the $^{212}$Bi and $^{214}$Bi background.

The construction of a medium-size BiPo-3 detector with a 3.25~m$^2$ sensitive surface and using the same techniques developed in the BiPo-1 prototype has been initiated. The goal of the BiPo-3 detector is to measure the first double beta decay source foils of the SuperNEMO demonstrator in the year 2011 with a sensitivity of $\mathcal{A}$($^{208}$Tl)~$<$~3-4~$\mu$Bq/kg (90\% C.L.) with a six month measurement. A first prototype has been running in the LSM since June 2010 and the backgrounds seems to have been reduced compared to BiPo-1 and the $^{214}$Bi measurement seems possible with BiPo-3. Another prototype with the full improvements for the BiPo-3 detector will be installed in Canfranc Underground Laboratory in the beginning of 2011 to confirm these preliminary results.

%%%%%%%%%%%%%%%%%%%%%%%%%%%%%%%%%%%%%%%%%%%%%%%%
%% BACKMATTER
%%%%%%%%%%%%%%%%%%%%%%%%%%%%%%%%%%%%%%%%%%%%%%%%

\begin{theacknowledgments}
The authors would like to thank the Modane Underground Laboratory staff for their technical support in running BiPo-1, and the IN2P3 Computing Center in Lyon for its software and computing support. This work was supported by the French Grant ANR-06-BLAN-0299 funded by the Agence Nationale de la Recherche, by the Spanish MICINN for the FPA2007-62833, FPA2008-03456 and FPA2006-12120-C03 contracts, part of which comes from FEDER funds, by the Russian RFBR 09-02-00737 Grant and by the UK STFC.
\end{theacknowledgments}

\bibliographystyle{aipproc}   % if natbib is available

\begin{thebibliography}{30}

\bibitem{nemo3a} R. Arnold et al., Nucl. Instr. and Meth. A 536 (2005) 79-122.
\bibitem{nemo3b} R. Arnold et al., Phys. Rev. Lett. 95 (2005) 182302.
\bibitem{nemo3c} R. Arnold et al., Nucl. Phys. A 765 (2006) 483-494.
\bibitem{nemo3d} R. Arnold et al., Nucl. Phys. A 781 (2007) 209-226.
\bibitem{nemo3e} R. Arnold et al., Nucl. Instr. and Meth. A 606 (2009) 449-465.
\bibitem{nemo3f} J. Argyriades et al., Phys. Rev. C 80 (2009) 032501.
\bibitem{nemo3g} V.I. Tretyak (on behalf of the NEMO-3 collaboration), AIP Conf. Proc. 1180 (2009) 135-139.

\bibitem{supernemoa} F. Piquemal (on behalf of the SuperNEMO collaboration), Phys. Atom. Nucl. 69 (2006) 2096-2010.
\bibitem{supernemob} R. Saakyan (on behalf of the SuperNEMO collaboration), J. Phys. Conf. Ser. 179 (2009) 012006.
\bibitem{breton} D. Breton et al., IEEE Trans. Nucl. Sci. 52 (2005) 2853.
\bibitem{lhcb} R. Cizeron et al., LHCb Calorimeter Technical Note LHCB 2000-52.
\bibitem{bipo1} SuperNEMO collaboration, Results of the BiPo-1 prototype for radiopurity measurements for the SuperNEMO double beta decay source foils, Nucl. Instr. and Meth. A 622 (2010) 120-128
\bibitem{bongrand} M. Bongrand (on behalf of the SuperNEMO collaboration), AIP Conf. Proc. 897 (2007) 14.
\bibitem{po212} Nuclear Data Sheets for A=212 104 (2005) 427.
\bibitem{borexino} G. Alimonti et al., Astropart. Phys. 8 (1998) 141-157.

\end{thebibliography}

\end{document}